\pdfoutput=1
\documentclass[a4paper,11pt]{article}
\usepackage{jheppub} % for details on the use of the package, please see the JINST-author-manual
\usepackage{lineno}
\usepackage{amsmath,amssymb,amsfonts,amsthm}
\usepackage{physics}
\usepackage[table]{xcolor}
\usepackage{multirow}
\usepackage{graphicx}% Include figure files
\usepackage{dcolumn}% Align table columns on decimal point
\usepackage{bm}% bold math
\usepackage{hyperref}% add hypertext capabilities
\usepackage[capitalise]{cleveref}

\usepackage{siunitx}
\definecolor{red}{rgb}{0.6,.0706,.1373}
\definecolor{blue}{rgb}{0,0.396,0.741}
\colorlet{blueRef}{blue!90!black}
\colorlet{blueLink}{blue!90!black}
\hypersetup{
	colorlinks, 
	bookmarksopen, 
	bookmarksnumbered,
	citecolor=blueRef, 		%color of links to bibliography
	linkcolor=blueLink,		%color of internal links
	urlcolor=blueLink,		%color of external links 
}

\newcommand{\U}{\mathrm{U}}
\newcommand{\SU}{\mathrm{SU}}

\newcommand{\LL}{\mathrm{L}}
\newcommand{\RR}{\mathrm{R}}

\newcommand{\diag}{\mathop{\mathrm{diag}}}

\newcommand{\eminus}{\vcenter{\hbox{\scalebox{0.6}[1]{$ - $}}}}	%Narrow minus signed (for e.g. negative exponents)
\newcommand{\rep}[1]{\mathbf{#1}}

\newcommand{\sscript}[1]{{\scriptscriptstyle \mathrm{#1}}}
\renewcommand{\L}{\mathcal{L}}

\title{Flavor Hierarchies From SU(2) Flavor and Quark--Lepton Unification}

\author[a]{Admir Greljo,\note{admir.greljo@unibas.ch}}%
\author[b]{Anders Eller Thomsen,\note{thomsen@itp.unibe.ch}}%
\author[a]{Hector Tiblom\note{hector.tiblom@unibas.ch}}
\affiliation[a]{%
 Department of Physics, University of Basel,\\Klingelbergstrasse 82, CH-4056 Basel, Switzerland}%
\affiliation[b]{%
 Albert Einstein Center for Fundamental Physics, Institute for Theoretical Physics,\\University of Bern, CH-3012 Bern, Switzerland}%
\date{\today}% It is always \today, today,
\abstract{In our recent attempt to explain flavor hierarchies~\cite{Greljo:2023bix}, a gauged SU(2) flavor symmetry acting on left-handed fermions provides a ground to introduce three independent rank-one contributions to the Yukawa matrices: a renormalizable one for the third family, a mass-suppressed one for the second family, and an additional loop-suppressed factor for the first family. Here, we demonstrate how minimal quark-lepton unification à la Pati-Salam, relating down-quarks to charged leptons, can significantly improve this mechanism. We construct and thoroughly analyze a renormalizable model, performing a comprehensive one-loop matching calculation that reveals how all flavor hierarchies emerge from a single ratio of two scales. The first signatures may appear in the upcoming charged lepton flavor violation experiments.}

\begin{document}
\maketitle
\flushbottom

%%%%%%%%%%%%%%%%%%
\section{Introduction}
\label{sec:intro}
%%%%%%

The peculiar pattern of fermion masses and their mixing under electroweak (EW) interactions stands as a riddle at the heart of the Standard Model (SM), known as \emph{the Flavor Puzzle}. All the flavor of the SM stems from Yukawa couplings of the three generations of fermions with the single Higgs field. Contrary to naive expectations, there are large hierarchies between the size of the coupling parameters, although no power counting is suggested by the model. Observations indicate a consistent mass hierarchy of approximately two orders of magnitude between consecutive generations of all electrically charged fermions: up quarks, down quarks, and charged leptons. Also, the misalignment between the up- and down-quark Yukawa matrices, as captured by the Cabibbo–Kobayashi–Maskawa (CKM) matrix~\cite{Cabibbo:1963yz, Kobayashi:1973fv}, is characterized by hierarchies between diagonal and off-diagonal elements spanning three orders of magnitude in total. 

The assumption of an underlying Beyond the Standard Model (BSM) explanation for these hierarchies may provide a valuable hint as to the nature of new physics, and there has been no lack of effort in this direction; Refs.~\cite{Froggatt:1978nt, Weinberg:1972ws, Barr:1990td, Babu:1990hu, Kaplan:1991dc, Leurer:1992wg, Leurer:1993gy, Kaplan:1993ej, Barbieri:1994cx, Barbieri:1995uv, Barbieri:1996ae, Barbieri:1996ww, Barbieri:1999pe, Randall:1999ee, Arkani-Hamed:1999ylh, King:2003rf, Grinstein:2010ve, Feruglio:2015jfa, Calibbi:2016hwq, Ema:2016ops, Panico:2016ull, Bordone:2017bld, Greljo:2018tuh, Linster:2018avp, Allanach:2018lvl, Alonso:2018bcg, Greljo:2019xan, Smolkovic:2019jow, Baur:2019kwi, Fedele:2020fvh, Nilles:2020nnc, King:2020qaj, Baker:2020vkh, Babu:2020tnf, Feruglio:2021dte, Altmannshofer:2021qwx, Davighi:2022fer, Cornella:2023zme, Asadi:2023ucx, Davighi:2023evx, Davighi:2023iks, Barbieri:2023qpf, Fuentes-Martin:2024fpx, Altmannshofer:2022aml} constitute a representative selection of early and recent examples. The goal of the flavor model building is to explain the many hierarchies observed in the fermion masses and mixings from a few (or no) ultraviolet (UV) hierarchies. We would judge the success of a mechanism based on how simple its UV realizations are and how well it produces the SM flavor pattern.

In physics, structures resulting from symmetries are widespread. Indeed, a compelling idea put forward very early on is that flavor hierarchies are a consequence of (accidental or gauged) flavor symmetries~\cite{Froggatt:1978nt}.  Relatively simple breaking patterns may be obtained from a horizontal $ \U(2) $ symmetry discriminating between the third and light generations~\cite{Babu:1990hu, Barbieri:1995uv}. It was recently observed~\cite{Greljo:2023bix, Antusch:2023shi} that imposing a single $\U(2)$ symmetry, which charges only one chirality of leptons and quarks respectively, still results in an accidental $ \U(2)^5 $~\cite{Feldmann:2008ja, Kagan:2009bn, Barbieri:2011ci, Barbieri:2012uh, Blankenburg:2012nx, Fuentes-Martin:2019mun, Faroughy:2020ina, Greljo:2022cah} of the SM Yukawa couplings in agreement with observations, allowing for a remarkably simple UV realization of the underlying mechanism. The key premise is to charge only the EW doublet on the quark side, which is involved in both the up and down Yukawa interactions. This minimal assignment generates the desired mass hierarchies and, at the same time, perturbative left-handed mixings entering the CKM matrix.

\begin{figure}[t]
    \centering
    \includegraphics[trim=3cm 21cm 3cm 5cm,width=\textwidth]{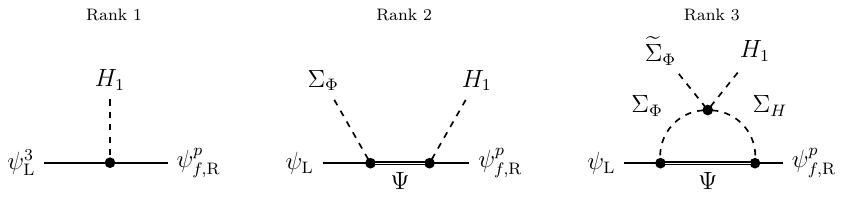}
    \caption{The mechanism for flavor hierarchies proposed in~\cite{Greljo:2023bix}. Three independent rank-one contributions to the Yukawa matrices: a renormalizable one for the third family (\textit{left}), a mass-suppressed one for the second family (\textit{center}), and an additionally loop-suppressed one for the first family (\textit{right}). The fields shown in the representative (and leading) Feynman diagrams are listed in Table~\ref{tab:matter_fields}.}
    \label{fig:diag}
\end{figure}

The model of~\cite{Greljo:2023bix} is based on an $ \SU(2)_{q+\ell} $ gauge symmetry under which the two light generations of left-handed quarks and leptons each form a doublet. This provides a basis to introduce three independent rank-one contributions to the Yukawa matrices, as shown in \cref{fig:diag}.  To begin with, the Higgs field can directly couple (and give mass) only to the third-generation fermions at the renormalizable level. Suppressed Higgs couplings with light generations are generated through the introduction of a symmetry-breaking scalar doublet, whose vacuum expectation value (VEV) breaks $ \SU(2)_{q+\ell} $. In particular, the inclusion of heavy vector-like quark and lepton fields generates at the tree level an effective dimension-5 operator between the flavor-breaking scalar, the Higgs, and light fermions; see \cref{fig:diag} (center). When a single copy of vector-like fermions (VLFs) is present, the corresponding rank of the Yukawa matrices is only raised by one unit, generating small non-zero Yukawa couplings for the second generation only.

In the final step, the effective Yukawa couplings for the first generation are loop-generated, reusing the same vector-like fermions; see \cref{fig:diag} (right). A clever construction that does not generate any additional unsuppressed dimension-5 operators (which might cause degeneracy between the light generation masses) is to complete the loops with new scalar leptoquark fields, such that a vector-like lepton (quark) enters that diagrams for the effective quark (lepton) Yukawa couplings. Technically, the dominant effect comes from the renormalization group mixing from a tree-level operator obtained after integrating out vector-like fermions solely. Remarkably, as long as the leptoquarks are lighter than the vector-like fermions, the hierarchy between the first two families is approximately a loop factor with a slight (logarithmic) dependence on the mass ratio.

The model proposed in Ref.~\cite{Greljo:2023bix} manages to explain the six mass hierarchies between consecutive generations of fermions in addition to the three hierarchies between the CKM matrix elements. All of this is in terms of two small input parameters $\epsilon_{q}$ and $\epsilon_\ell$, the ratios of the flavor symmetry-breaking scale to the two vector-like fermion masses. The similarity of the down quark and charged lepton mass hierarchies imply $\epsilon_{q} \approx \epsilon_\ell$, which, in this model, requires the two independent vector-like fermion masses to coincide. 
The model involves a single new gauge symmetry, SM$\times \SU(2)_{q+\ell}$, and six new matter fields. At first, this might seem unnecessarily convoluted compared to some mechanisms to explain the flavor puzzle, but when judging realizations in UV complete models, it compares very favorably. In fact, it admits further simplification and unification, as we have detailed in this paper. 

The field content of the $ \SU(2)_{q+\ell} $ model~\cite{Greljo:2023bix} is very suggestive of a \textit{minimal} quark--lepton unification in the manner of Pati--Salam~\cite{Pati:1974yy} embedding color, lepton number and baryon number into an $ \SU(4) $ gauge group, but without introducing $ \SU(2)_\RR $~\cite{Smirnov:1995jq,FileviezPerez:2013zmv}. The embedding of the gauge groups collects both vector-like fermions into a single vector-like fourplet, thereby explaining the coincidence of scales between the two, \textit{predicting} $\epsilon_q = \epsilon_\ell$. Additionally, the four scalar fields introduced in the original model are embedded into just two scalar multiplets. Quark--lepton unification thus provides a simple extension to the model, reducing the BSM matter content to four new fields and further unifying the SM chiral fermions.

The goal of the present paper is to embed the model of Ref.~\cite{Greljo:2023bix} into the $ \SU(4) $ gauge group, unifying quarks and leptons. We will demonstrate how the SM flavor structure emerges from a \textit{single} hierarchy in the UV. A full-fledged renormalizable model is defined in \cref{sec:model}, while the matching calculation of the effective SM Yukawa couplings is presented in \cref{sec:matching}. In \cref{sec:numerics}, we show how the SM masses are recovered from a probability distribution by varying order-one parameters. Finally, in \cref{sec:pheno}, we discuss $ K_L \to \mu e$ and muon conversion on heavy nuclei as the main phenomenological constraints on the model, pointing to a symmetry-breaking scale above a few PeV. Future sensitivity in upgraded muon conversion experiments provides an opportunity for discovery if the flavor-breaking scale is low enough. Conclusions are left for \cref{sec:conc} while the details of the scalar potential are relegated to the Appendix.

%%%%%%%
\section{The model: Quark--lepton unification with gauged $\SU(2)$ flavor}
\label{sec:model}
%%%%%%%%
As a starting point for our model, we embed $\SU(3)_c$ and $\U(1)_{\rm B-L}$ into an $ \SU(4) $ gauge group in the manner of Pati and Salam~\cite{Pati:1974yy}, identifying leptons with a ``fourth color". The SM hypercharge emerges from the product of $ \SU(4) $ with a new Abelian group $ \U(1)_\RR $ acting on the right-handed SM fermions. At low scales, we expect to find the SM symmetry embedded as $ \SU(3)_c \times \U(1)_Y \subseteq \SU(4) \times \U(1)_\RR $. This construction corresponds to the minimal quark--lepton unification~\cite{Smirnov:1995jq, FileviezPerez:2013zmv, FileviezPerez:2023rxn} and can manifest at energies as low as the PeV-scale. The $ \SU(4) \times \U(1)_\RR $ gauge group is further augmented with the $ \SU(2)_\LL $ of the SM in addition to the $ \SU(2)_{q+\ell} $ that acts on the two light generations of the left-handed SM fermions~\cite{Babu:1990hu, Greljo:2023bix}. The latter group is the origin of the flavor structure of the model and is responsible for producing the SM mass hierarchies. Thus, the UV gauge group is given by 
    \begin{equation}
        G_{\sscript{UV} }  =  \SU(4) \times \SU(2)_\LL \times \U(1)_\RR \times \SU(2)_{q+\ell}.
    \end{equation}
The full matter content of our theory is shown in Table~\ref{tab:matter_fields}. The SM fermions fit into the four irreducible representations under $ G_{\sscript{UV} } $: $ \psi_\LL = (q_\LL,\, \ell_\LL) $ for the two light generation left-handed fermions which form an $ \SU(2)_{q+\ell} $ doublet; $ \psi^3_\LL = (q^3_\LL,\, \ell^3_\LL) $ for the third generation left-handed fermions; three copies of $ \psi^p_{d,\RR} = (d^p_\RR,\, e^p_\RR) $ make up all generations for the down-type quarks and charged leptons; and three copies of $ \psi^p_{u, \RR} = (u^p_\RR,\, \nu^p_\RR) $, which contain the three right-handed up-type quarks. To complete $ \psi^p_{u, \RR} $, one must also include three right-handed neutrinos in the SM field content. We postpone the discussion of neutrino masses and mixings to \cref{sec:neutrino}. In order to generate the correct flavor structure, we additionally include a single vector-like fermion $ \Psi_{\LL,\RR} $, which transforms as the third-generation left-handed fermions.

\begin{table}[t]
    \centering
    \begin{tabular}{|c|c|c|c|c|}\hline
    \rowcolor{black!15}[\tabcolsep] Field & $\mathrm{SU}(4)$ & $\mathrm{SU}(2)_\mathrm{L}$& $\mathrm{U}(1)_\mathrm{R}$& $\mathrm{SU}(2)_{q+\ell}$ \\\hline
    $\psi_\mathrm{L}$ & $\mathbf{4}$ & $\mathbf{2}$ & $\phantom{\eminus}0\phantom{\eminus}$ & $\mathbf{2}$ \\
    $\psi_\mathrm{L}^3$ & $\mathbf{4}$ & $\mathbf{2}$ & $\phantom{\eminus}0\phantom{\eminus}$ & $\mathbf{1}$ \\
    $\psi_{u,\RR}^p$ & $\mathbf{4}$ & $\mathbf{1}$ & $\phantom{\eminus}1/2\phantom{\eminus}$ & $\mathbf{1}$ \\
    $\psi_{d,\RR}^p$ & $\mathbf{4}$ & $\mathbf{1}$ & $\eminus 1/2\phantom{\eminus}$ & $\mathbf{1}$ \\\hline
    $\Psi_\mathrm{L,R}$ & $\mathbf{4}$ & $\mathbf{2}$ & $\phantom{\eminus}0\phantom{\eminus}$ & $\mathbf{1}$ \\\hline\hline
    $\chi$ & $\mathbf{4}$ & $\mathbf{1}$ & $\phantom{\eminus}1/2\phantom{\eminus}$ & $\mathbf{1}$ \\
    $H_1$ & $\mathbf{1}$ & $\mathbf{2}$ & $\phantom{\eminus}1/2\phantom{\eminus}$ & $\mathbf{1}$ \\
    $\Sigma_\mathrm{H}$ & $\mathbf{15}$ & $\mathbf{2}$ & $\phantom{\eminus}1/2\phantom{\eminus}$ & $\mathbf{1}$ \\
    $\Sigma_\mathrm{\Phi}$ & $\mathbf{15}$ & $\mathbf{1}$ & $\phantom{\eminus}0\phantom{\eminus}$ & $\mathbf{2}$ \\\hline
    \end{tabular}
    \caption{The matter field content of the model and their representations under the gauge group. The first four rows are chiral fermions, $ \Psi $ is a vector-like fermion, while the remaining fields are scalars. There are three copies of the right-handed chiral fermions ($p=1,2,3$), but other fields do not replicate. }
    \label{tab:matter_fields}
\end{table}

The flavor hierarchies of~\cite{Greljo:2023bix} were produced through the use of five scalar fields. These fields can be embedded into just two irreducible representations of $ G_{\sscript{UV} } $: $\Sigma_H$ and $\Sigma_\Phi$. Two additional scalars are required, so our model comprises four scalar fields: $ \chi $, which is used to break $ \SU(4) \times \U(1)_\RR $ to $ \SU(3)_c \times \U(1)_Y $; $H_1$, which is used to break the EW symmetry and provide fermion masses; $ \Sigma_H $, which contains both the leptoquarks that run in the loops as well as an additional Higgs that can split the masses between the quark and lepton sectors; and, lastly, $ \Sigma_\Phi $, which contains additional necessary leptoquarks and is used to break the flavor symmetry $ \SU(2)_{q+\ell} $. 
The scalar field $ \Sigma_\Phi $ contains an SM singlet component $ \Phi $, the VEV of which will break the flavor symmetry. At first glance, it seems like the inclusion of $ \Sigma_\Phi $ is not a minimal configuration since it is sufficient for $\chi$ and a $ \rep{2}_{q+\ell} $ doublet $\Phi$ to develop VEVs to break $ G_{\sscript{UV} } $ into the SM gauge group. However, to provide masses to the first generation as shown in~\cite{Greljo:2023bix}, we need the leptoquark $S$, which is a triplet under the color symmetry and a doublet under the flavor symmetry. $ \Sigma_\Phi $ contains both $S$ and $\Phi$ and is, therefore, the smallest representation that simultaneously ensures the breaking of the flavor symmetry and the masses for the first family. For the decomposition of the fields under $ G_\sscript{SM} \times \SU(2)_{q+\ell} $, see \cref{tab:component_fields}.

\begin{figure}
    \centering
    \includegraphics[width=.8\textwidth]{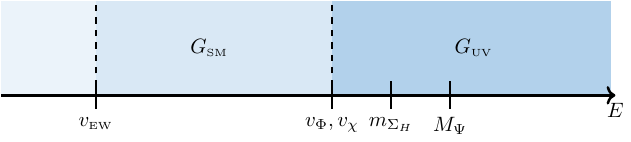}
    \caption{A benchmark scenario for the mass thresholds of the model on a logarithmic scale.}
    \label{fig:scales}
\end{figure}

As per usual in non-supersymmetric BSM models with new scalar states, the scalar potential in our model is rather complicated and contains in excess of 60 free parameters; we refer the reader to \cref{app:A} for the full potential. Of these there are only four parameters with the mass dimension, which control the scales on the model (see \cref{fig:scales}). The rest are marginal couplings and are assumed to be of $\mathcal{O}(1)$, apart from complying with the vacuum stability and perturbative unitarity bounds. With the full potential in hand, we solved the tadpole equations and verified that our VEV configuration does indeed minimize the potential. Scanning through the parameter space, we easily find suitable points that produce condensates for $\chi$, $\Phi$, and one of the Higgses while still ensuring all other scalar fields obtain positive masses.

\begin{table}[t]
    \centering
    \begin{tabular}{|c|c|c|c|c| c|}\hline
    \rowcolor{black!15}[\tabcolsep] UV Field &  Component & $\mathrm{SU}(3)$ & $\mathrm{SU}(2)_\LL$ & $\mathrm{U}(1)_Y$& $\mathrm{SU}(2)_{q+\ell}$ \\ \hline
    \multirow{4}{*}{$\Sigma_\mathrm{H}$} & $\Theta_H$ & $\rep{8}$ & $\rep{2}$ & $\phantom{\eminus}0\phantom{\eminus}$ & $ \rep{1}$\\ 
    & $ R_u $ & $\rep{3}$ & $\rep{2}$ & $ \phantom{\eminus}7/6\phantom{\eminus} $ & $ \rep{1} $\\
    & $ R_d $ & $\rep{3}$ & $\rep{2}$ & $ \phantom{\eminus}1/6\phantom{\eminus} $ & $ \rep{1} $\\
    & $H_2$ & $\rep{1}$ & $\rep{2}$ & $\phantom{\eminus}1/2\phantom{\eminus}$ & $ \rep{1} $\\ \hline
    \multirow{4}{*}{$\Sigma_\mathrm{\Phi}$} & $\Theta_\Phi$ & $\rep{8}$ & $\rep{1}$ & $\phantom{\eminus}0\phantom{\eminus}$ & $ \rep{2}$\\
    & $S_1$ & $\rep{3}$ & $\rep{1}$ & $\eminus 1/3 \phantom{\eminus}$ & $ \rep{2}$\\
    & $S_2$ & $\rep{3}$ & $\rep{1}$ & $\eminus 1/3 \phantom{\eminus}$ & $ \rep{2}$\\
    & $\Phi$ & $\rep{1}$ & $\rep{1}$ & $\phantom{\eminus}0\phantom{\eminus}$ & $ \rep{2}$\\ \hline
    \end{tabular}
    \caption{Decomposition of the scalar fields under $ G_\sscript{SM} \times \SU(2)_{q+\ell} $.}
    \label{tab:component_fields}
\end{table}

It is rather simple to organize the symmetry-breaking pattern: 
A non-zero VEV of $ \chi $ is responsible for the breaking $ G_{\sscript{UV} } $ to $G_\sscript{SM} \times \SU(2)_{q+\ell}$. With an $ \SU(4) $ rotation, the VEV can always be arranged such that 
\begin{equation}
    \ev{\chi}=\mqty(0\\0\\0\\v_\chi).
\end{equation}
In the broken phase, the adjoint scalar fields then decompose as in \cref{tab:component_fields}. In matrix form
\begin{equation}
    \Sigma_H = \mqty(\Theta_H&R_u\\\widetilde{R}_d&0)+T_{15} H_2\qc \Sigma_\Phi = \mqty(\Theta_\Phi&S_1\\\widetilde{S}_2&0)+T_{15} \Phi,
\end{equation}
where for any $\SU(2)$ doublet we use the notation $\widetilde{F}^i=\varepsilon^{ij}F_j^*$, and $T_{15}$ is the diagonal generator of $\SU(4)$ in the fundamental representation,
\begin{equation}
    T_{15}=\frac{1}{2\sqrt{6}}\mqty(\dmat[\phantom{N}]{1,1,1,-3}).
\end{equation}
In \cite{Greljo:2023bix}, the masses for the first family are generated through the use of three leptoquarks: $R_u,R_d,S$. These leptoquarks are embedded into the adjoint fields $\Sigma_H,\Sigma_\Phi$, with the only difference being that the latter contains two leptoquarks $S_1,S_2$ with the quantum numbers of $S$. To break the flavor symmetry in the same way, $\Phi$ develops a VEV according to
\begin{equation}
    \ev{\Phi}=\mqty(0\\v_\Phi),\quad  \rm{where} \quad\langle \widetilde \Phi \rangle=\mqty(v_\Phi \\ 0).
\end{equation}
The placement of the VEV picks out light flavors. 

The flavor hierarchies are controlled by $\varepsilon=v_\Phi / M_\Psi \sim 10^{-2}$, where $M_\Psi$ is the VLF mass. There is freedom in choosing $v_\chi$. However, simple naturalness considerations of the scalar potential suggest $v_\chi \sim v_\Phi$ since the portal operators generically transmit VEVs into dimensionful couplings among the remaining scalars.\footnote{Scalar leptoquarks entering the loop diagrams for the first family masses (\cref{fig:diag} right) must have a mass no larger than $\mathcal{O}(M_\Psi)$. Otherwise, the hierarchy between the first two families becomes excessively large. For $\mathcal{O}(1)$ scalar quartics, this condition implies $v_\chi \lesssim M_\Psi$. Thus, without tuning, the $\SU(4)$ breaking can not be decoupled from the flavor breaking when assuming all dimensionless couplings are $\mathcal{O}(1)$.}  Thus, for simplicity, we will assume that $ \chi $ and $ \Phi $ develop VEVs at the same scale, as expected in the absence of tuning. As we will see in \cref{sec:pheno}, current experimental data indicates that the symmetry-breaking should occur at the PeV scale or higher. 

Indeed, it is an experimental necessity that the EW symmetry breaking occurs at a much lower scale than the breaking of $ G_\sscript{UV} $ to $ G_\sscript{SM} $ (see Section~\ref{sec:pheno}). Admittedly, a tuning of the scalar potential is required in order to get such a separation of scales of more than three orders of magnitude. This is an irreducible problem known to all BSM models with fundamental scalars in the absence of a protection mechanism such as supersymmetry, and we will not resolve this issue here. After the gauge symmetry breaks to $ G_\sscript{SM}$, the two Higgs fields $(H_1, H_2)$ get an effective mass matrix 
    \begin{equation}
    M^2_H = \begin{pmatrix}
        m_{H_1}^{\prime 2} & - \frac{\left(\lambda _{51}+\lambda _{52}\right) v_{\Phi }^2+3 \lambda _5 v_{\chi }^2}{2 \sqrt{6}} \\
        - \frac{\left(\lambda _{51}+\lambda _{52}\right) v_{\Phi }^2+3 \lambda _5 v_{\chi }^2}{2 \sqrt{6}} & m_{\Sigma_H}^{\prime 2}
    \end{pmatrix},
    \end{equation}
mixing the two components. The diagonal components receive contributions from various quartic interactions with $ \langle \Sigma_\Phi \rangle $ and $ \langle \chi \rangle $ insertions, which, for our purposes, are simply absorbed into the mass terms. A realistic limit with an SM-like Higgs boson is obtained when $m_{H_1}^{\prime 2} \ll \frac{\left(\lambda _{51}+\lambda _{52}\right) v_{\Phi }^2+3 \lambda _5 v_{\chi }^2}{2 \sqrt{6}} \lesssim m_{\Sigma_H}^{\prime 2} $. This ensures a small but non-zero mixing between the two states, where we consider a \textit{single} tuning of $ m_{H_1}^{2} $ to ensure that the state consisting mostly of $ H_1 $ gets a small, negative mass, thereby developing an EW symmetry-breaking VEV. In this scenario, the ratio between VEVs of $ H_1 $ and $ H_2 $ is given by
\begin{equation}
    \tan \beta\equiv  \frac{v_2}{v_1} = -\frac{\left(\lambda _{51}+\lambda _{52}\right) v_{\Phi }^2+3 \lambda _5 v_{\chi }^2}{2 \sqrt{6} \,m_{\Sigma_H}^{\prime 2} }\,,
\end{equation} 
where $v_1^2+v_2^2=v_\sscript{EW}^2 = (\SI{174}{GeV})^2$. A modest order of magnitude difference in the scales $v_{\Phi, \chi}$ and $m_{\Sigma_H}$, easily gives $ \tan \beta \lesssim 0.01 $ without any additional tuning. A small $\tan \beta$ is particularly interesting for phenomenology, as will be discussed in \cref{sec:numerics}, suggesting $v_{\Phi, \chi} < m'_{\Sigma_H} < M_\Psi$. A rough sketch of a realistic benchmark for the various scales is shown in \cref{fig:scales}.

%%%%%%%%%%%
\section{Matching to the SM Yukawas}
\label{sec:matching}
%%%%%%%%%%%

The fermion masses are generated through a combination of renormalizable and dimension-5 effective operators produced at both tree and one-loop levels, starting from the UV Lagrangian and integrating out heavy fields. We will examine these contributions individually and establish their role in determining the rank of the mass matrices.

%%%%
\subsection{Rank 1}
%%%%

The symmetries of the model only allow the third generation to acquire masses at the renormalizable level. The relevant Yukawa interactions are 
\begin{equation}
    \mathcal{L}_\sscript{UV}\supset-x_\mathrm{u}^p\overline{\psi}_\mathrm{L}^3\widetilde{H}_1\psi_{u,\mathrm{R}}^p-X_\mathrm{u}^p\overline{\psi}_\mathrm{L}^3\widetilde{\Sigma}_H\psi_{u,\mathrm{R}}^p
    -x_\mathrm{d}^p\overline{\psi}_\mathrm{L}^3H_1\psi_{d,\mathrm{R}}^p-X_\mathrm{d}^p\overline{\psi}_\mathrm{L}^3\Sigma_H\psi_{d,\mathrm{R}}^p+\text{h.c.}.
\end{equation}
We denote the couplings to the singlet Higgs field $H_1$ with lowercase letters and the couplings to the adjoint Higgs field $\Sigma_H$ with uppercase letters. When the EW symmetry is broken by the Higgs fields developing VEVs, the resulting SM Yukawa matrices $Y_f^{pr}$ ($p,r=1,2,3$) will depend on a fixed linear combination of the vectors $x_{u(d)}^p$ and $X_{u(d)}^p$. Importantly, these will contribute only to the third rows, resulting in the Yukawa matrices having rank 1.

%%%%
\subsection{Rank 2}
%%%%

To raise the rank of the Yukawa matrices and provide the second generation with masses, we make use of the renormalizable interactions between the VLF $\Psi$ and the scalar fields. After integrating out $\Psi$, we will be left with dimension-5 operators that generate masses for the second generation. The relevant part of the Lagrangian for the matching calculation is 
\begin{alignat}{2}
    \mathcal{L}_\sscript{UV}\supset-&y_\mathrm{u}^p\overline{\Psi}_\mathrm{L}\widetilde{H}_1\psi_{u,\mathrm{R}}^p&-&Y_\mathrm{u}^p\overline{\Psi}_\mathrm{L}\widetilde{\Sigma}_H\psi_{u,\mathrm{R}}^p\nonumber\\
    -&y_\mathrm{d}^p\overline{\Psi}_\mathrm{L}H_1\psi_{d,\mathrm{R}}^p&-&Y_\mathrm{d}^p\overline{\Psi}_\mathrm{L}\Sigma_H\psi_{d,\mathrm{R}}^p\nonumber\\
    -&Y_\mathrm{\Phi}\overline{\psi}_\mathrm{L}\Sigma_\Phi\Psi_\mathrm{R}&-&\widetilde{Y}_\Phi\overline{\psi}_\mathrm{L}\widetilde{\Sigma}_\Phi\Psi_\mathrm{R}+\text{h.c.}\,.\label{eq:veccoup}
\end{alignat}
By redefining fields, we have chosen a basis with no mass mixing between $\Psi_\mathrm{L}$ and $\psi^3_\mathrm{L}$. In addition, note that $\Sigma_\Phi$ and $\widetilde{\Sigma}_\Phi$ share quantum numbers. Thus, the kinetic term for $\Sigma_\Phi$ exhibits an accidental $\SU(2)$ symmetry. We exploit this symmetry by choosing a basis where $\widetilde{Y}_\Phi=0$, ensuring that the first generation of fermions remains massless at the tree level. Integrating out $\Psi$ at the tree level, we are left with the dimension-5 operators
\begin{alignat}{2}
    \mathcal{L}_\sscript{EFT}\supset\frac{Y_\Phi}{M_\Psi}\Big(&y_\mathrm{u}^p\overline{\psi}_\mathrm{L}\Sigma_\Phi\widetilde{H}_1\psi_{u,\mathrm{R}}^p&+&Y_\mathrm{u}^p\overline{\psi}_\mathrm{L}\Sigma_\Phi\widetilde{\Sigma}_H\psi_{u,\mathrm{R}}^p\nonumber\\
    +&y_\mathrm{d}^p\overline{\psi}_\mathrm{L}\Sigma_\Phi H_1\psi_{d,\mathrm{R}}^p&+&Y_\mathrm{d}^p\overline{\psi}_\mathrm{L}\Sigma_\Phi\Sigma_H\psi_{d,\mathrm{R}}^p+\text{h.c.}\Big).
\end{alignat}

%%%%%%
\subsection{Rank 3}
%%%%%%

Already at the one-loop level, the rank of the mass matrices is increased to 3. As a result, the first family becomes massive, and the loop factor nicely explains the hierarchy between the two light families. Even though the same couplings from Eq.~\eqref{eq:veccoup} are involved in the loop diagrams (see \cref{fig:diag} right), the leptoquark contributions, together with the independent scalar quartics, provide enough freedom to raise the rank and correctly fit the first-family parameters.  We use \texttt{Matchete}~\cite{Fuentes-Martin:2022jrf} to compute the effective dimension-5 Yukawa operators at the one loop. Due to their length and complexity, we only list one of the representative terms which increases the rank:
\begin{align}
    \mathcal{L}_\sscript{EFT}\supset\frac{15}{32}\frac{1}{16\pi^2}\frac{1}{M_\Psi}\qty[\log \qty(\frac{M_{\Psi }^2}{\mu ^2})-1]Y_u^p Y_\Phi\lambda_{45}^*\Tr[\widetilde{\Sigma}_H\widetilde{\Sigma}_\Phi]\overline{\psi}_\mathrm{L}\psi_{u,\mathrm{R}}^p.
\end{align}
When the scalar fields develop VEVs, the leading expressions for the resulting mass matrices are rather simple and will be presented in full below.

%%%%%%%
\subsection{The mass matrices}
%%%%%%%

The fermion masses are generated after the Higgs doublets develop VEVs according to
\begin{equation}
    \ev{H_i}=\mqty(0\\v_i).
\end{equation}
$H_1$ will contribute equally to the quark and lepton sectors since it is a singlet under $\SU(4)$, while $H_2$ will split the down-quark and charged-lepton masses since it is embedded in the adjoint field $\Sigma_H$. 

Utilizing the remaining freedom to rotate the right-handed fields, $\U(3)_{\psi_{u,\RR}} \times \U(3)_{\psi_{d,\RR}}$, we choose a basis where the tree-level quark Yukawa matrices are upper-triangular,\footnote{The conditions are $ 2 \sqrt{6} v_1 y_{f}^1+v_2 Y_{f}^1=0$ and $v_1 x_{f}^r+\frac{v_2 X_{f}^r}{2 \sqrt{6}}=0$ where $r=1,2$.
}
\begin{equation}
M_{f}^\sscript{Tree}=\mqty(
 0& 0 & 0 \\
 0& -\frac{v_{\Phi } Y_{\Phi } \qty(2 \sqrt{6} v_1 y_{f}^2+v_2 Y_{f}^2)}{24 M_{\Psi }} & -\frac{v_{\Phi } Y_{\Phi } \qty(2 \sqrt{6} v_1 y_{f}^3+v_2 Y_{f}^3)}{24 M_{\Psi }} \\
 0 & 0 & v_1 x_{f}^3+\frac{v_2 X_{f}^3}{2 \sqrt{6}} 
)\qc f\in\qty{u,d}.
\end{equation}
Including the one-loop results only for the first generation, where it makes the difference, we finally get 
\begin{equation}\label{eq:Mq}
M_{u}=\mqty(
 -b\qty(2 \sqrt{6} v_1 \lambda _{50}^*+v_2 \lambda_u^*)Y_u \\
 -a\qty(2 \sqrt{6} v_1 y_{u}+v_2 Y_{u}) \\
 v_1 x_{u}+\frac{v_2 X_{u}}{2 \sqrt{6}})\qc M_{d}=\mqty(
 b\qty(2 \sqrt{6} v_1 \lambda _{49}^*+v_2 \lambda_d^*)Y_d \\
 -a\qty(2 \sqrt{6} v_1 y_{d}+v_2 Y_{d}) \\
 v_1 x_{d}+\frac{v_2 X_{d}}{2 \sqrt{6}}).
\end{equation}
Here
\begin{equation}
    a=\frac{Y_\Phi}{24}\frac{v_\Phi}{M_\Psi}\qc b=\frac{1}{16\pi^2}\frac{Y_\Phi}{24}\frac{v_\Phi}{M_\Psi}\qty[\log \qty(\frac{M_{\Psi }^2}{\mu ^2})-1],
\end{equation}
are the suppression factors associated with the tree-level and one-loop operators, respectively, and
\begin{equation}
    \lambda_u=-11\lambda_{45}-\lambda_{46}+3\lambda_{47}+2 \lambda_{48}\qc
    \lambda_d=-11\lambda_{45}-\frac{391}{128}\lambda_{46}+ \lambda_{47}+2\lambda_{48}.
\end{equation}
Note that the quark mass matrices are upper-triangular with the hierarchical rows, which ensures small left-handed rotation matrices in agreement with the CKM matrix.

In our chosen basis, the charged lepton mass matrix $M_e$ is not upper-triangular due to its dependence on the same Yukawa couplings as the down-type quark mass matrix, i.e., $\U(3)_{\psi_{d,\RR}}$ is already used up to set the latter into a desired form. Therefore, 
\begin{equation}\label{eq:Me}
M_e=\resizebox{.83\hsize}{!}{$\mqty(
 -3b\qty(2\sqrt{6} v_1\lambda_{49}^*+v_2\lambda_e)Y_d^1 & -3b\qty(2\sqrt{6} v_1\lambda_{49}^*+v_2\lambda_e)Y_d^2 & -3b\qty(2\sqrt{6} v_1\lambda_{49}^*+v_2\lambda_e)Y_d^3 \\
 -12av_2 Y_d^1 & 3a\qty(2 \sqrt{6} v_1 y_d^2-3 v_2 Y_d^2) & 3a\qty(2 \sqrt{6} v_1 y_d^3-3 v_2 Y_d^3) \\
 -\sqrt{\frac{2}{3}} v_2 X_d^1 & -\sqrt{\frac{2}{3}} v_2 X_d^2 & v_1 x_d^3-\frac{1}{2} \sqrt{\frac{3}{2}} v_2 X_d^3)$},
\end{equation}
where
\begin{equation}
     \lambda_e=3\lambda_{45}+\frac{149}{128}\lambda_{46}-3 \lambda_{47}+2\lambda_{48}.
\end{equation}

%%%%%%%%%%%%%
\section{Producing the flavor hierarchies}
\label{sec:numerics}
%%%%%%%%%%%%%

Starting from the mass matrices in Eqs.~\eqref{eq:Mq} and \eqref{eq:Me}, in \cref{sec:pertdiag}, we will derive quark and charged-lepton masses and the CKM elements in terms of the UV parameters. Subsequently, in \cref{sec:numerics2}, we will conduct a detailed numerical scan of the UV parameter space to predict the SM flavor hierarchies. Finally, we will discuss the neutrino sector in \cref{sec:neutrino}.

%%%%%
\subsection{Perturbative diagonalization}
\label{sec:pertdiag}
%%%%%

Since the quark mass matrices are upper-triangular with hierarchical rows, we perturbatively diagonalize them to obtain the quark masses and the CKM matrix. The perturbative diagonalization is performed as described in the Appendix of Ref.~\cite{Greljo:2023bix}, and we report only the leading terms in $b\ll a\ll 1$ expansion. The singular value decomposition gives us the rotation matrices through
\begin{equation}
    M_{u(d)} = L_{u(d)}\widehat{M}_{u(d)}R_{u(d)}^\dagger,
\end{equation}
where $\widehat{M}_{u(d)}$ is diagonal. The upper-triangular structure of the quark mass matrices ensures that the off-diagonal elements of $R_{u(d)}$ are negligible, while the hierarchy between the rows ensures that the off-diagonal elements of $L_{u(d)}$ are small and hierarchical, satisfying
\begin{equation}
    \qty[L_{u(d)}]_{ij} \propto \frac{\qty[\widehat{M}_{u(d)}]_{ii}}{\qty[\widehat{M}_{u(d)}]_{jj}}\qc i\leq j.
\end{equation}
The CKM matrix is then determined by $V_\mathrm{CKM}=L_u^\dagger L_d$.
From the diagonal $\widehat{M}_{u(d)}$, we get the quark masses
\begin{alignat}{2}
    m_u &= -b\qty(2 \sqrt{6} v_1 \lambda _{50}^*+v_2 \lambda_u^*)Y_u^1\qc &m_d &= b\qty(2 \sqrt{6} v_1 \lambda _{49}^*+v_2 \lambda_d^*)Y_d^1, \\
    m_c &= -a\qty(2 \sqrt{6} v_1 y_{u}^2+v_2 Y_{u}^2)\qc &m_s &= -a\qty(2 \sqrt{6} v_1 y_{d}^2+v_2 Y_{d}^2), \\
    m_t &= v_1 x_{u}^3+\frac{v_2 X_{u}^3}{2 \sqrt{6}}\qc &m_b &= v_1 x_{d}^3+\frac{v_2 X_{d}^3}{2 \sqrt{6}}.
\end{alignat}
While using the full expression in the numerical scan, we omit writing it here due to its complexity and only note that it satisfies $\qty[V_\mathrm{CKM}]_{ii}=1$ at the leading order.

The left-handed rotations in the lepton sector are also small and hierarchical due to the hierarchies between the rows in \cref{eq:Me}. By contrast, the right-handed rotations are can be sizeable depending on the region of parameter space. Consider now the limit $\tan \beta \to 0$, in which $M_e$ becomes upper-triangular, causing the right-handed rotations to be negligible. In this limit it follows from \cref{eq:Mq,eq:Me} that the third rows of $M_d$ and $M_e$ become identical while the first two rows differ only by a multiplicative factor. In this limit, $m_b = m_\tau = v_1 x_d^3$. Empirically, the bottom-tau unification is remarkably successful at high energies, which is the primary motivation for considering this limit. However, obtaining the correct size necessitates restricting $ x_d^3 \sim \mathcal{O}(10^{-2}) $. As we will demonstrate in the numerical scan below, this is the only dimensionless parameter that must be somewhat small compared to $ \mathcal{O}(1)$ expectations. In fact, as long as $\tan \beta \lesssim y_{b,\tau } \sim 10^{-2}$, only a single parameter needs to be tuned. For $\tan \beta \sim 1$, four parameters must be restricted to magnitudes of $10^{-2}$, see \cref{eq:Me}. Finally, the $\tan \beta \to \infty$ limit predicts a wrong mass ratio $m_b/m_\tau=1/3$ and is ruled out. In the numerical studies, we will focus on $\tan \beta \lesssim 10^{-2}$.

%%%%%%
\subsection{Numerical scan}
\label{sec:numerics2}
%%%%%%

\begin{figure}[t]
    \centering
    \includegraphics{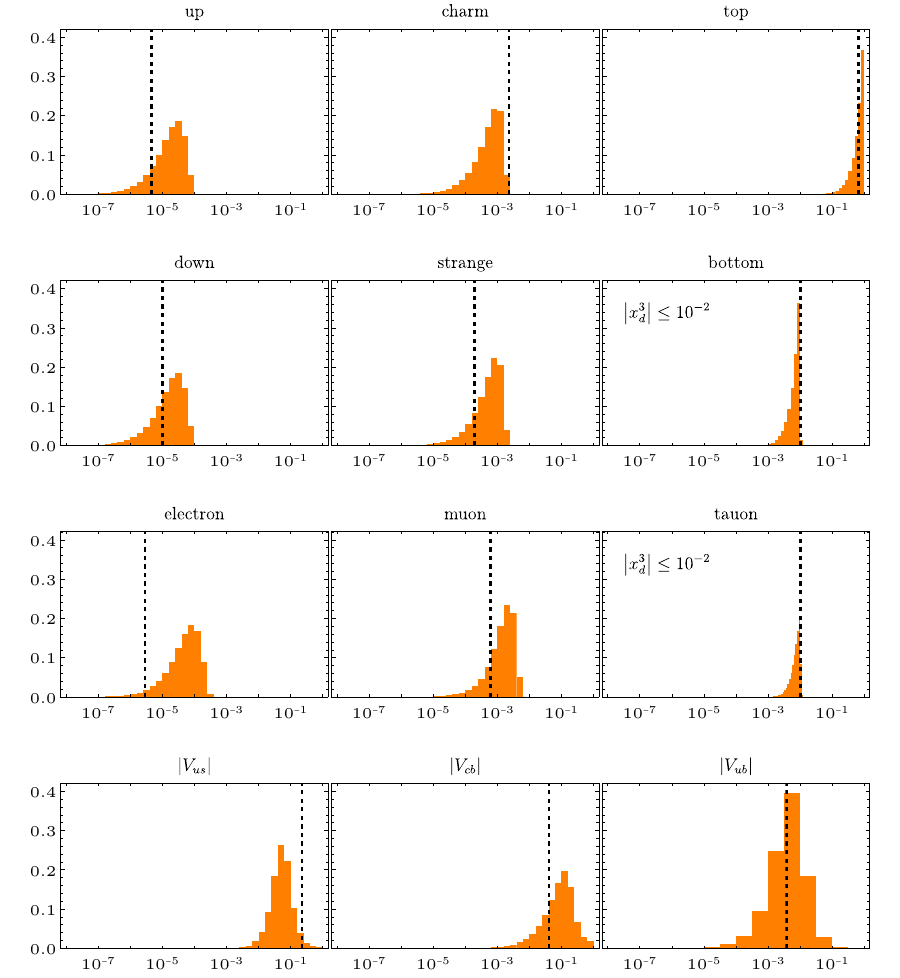}
    \caption{Histogram showing the probability of obtaining the correct order of magnitude for the SM flavor parameters when the UV parameters take on random numbers drawn from a flat distribution with the magnitude $\leq 1$. The black lines display the running SM values at the renormalization scale \SI{1}{PeV}. See Section~\ref{sec:numerics2} for details.}
    \label{fig:hist}
\end{figure}

We randomly generate one million parameter points from a flat distribution, applying the following constraints on the dimensionless couplings in the UV Lagrangian: $-1\leq Y_\Phi\leq 1$, $\abs{x_d^3}\leq 10^{\eminus 2}$,
and all other contributing Yukawa and scalar couplings are allowed to be complex, with their magnitudes constrained to $\leq1$. We take the symmetry breaking and the renormalization scales equal, $v_\Phi = \mu = \SI{1}{PeV}$, and fix the ratio between the VLF mass and the symmetry breaking scale to $\varepsilon = v_\Phi/M_\Psi = 10^{\eminus 2}$. Additionally, we set $ \tan \beta =0.001 $, but we have verified that the results remain qualitatively unchanged as long as $\tan \beta \lesssim 10^{-2}$. 

In \cref{fig:hist}, we plot the predicted histograms for the SM flavor parameters. The first three rows display the values of the Yukawa couplings, $ y_i = m_i/v_\sscript{EW} $, while the fourth row shows the magnitudes of the CKM elements. The black solid lines represent the singular values of the running SM Yukawa couplings at the renormalization scale $\mu$ as given in~\cite{Greljo:2023bix}, and the CKM matrix elements, which do not run significantly, as provided by the PDG~\cite{Workman:2022ynf}. These distributions illustrate the overall satisfactory agreement between the predicted and observed parameters. Somewhat ad-hoc restriction of a single parameter $\abs{x_d^3}\leq 10^{\eminus 2}$ is critical; otherwise, for $\abs{x_d^3}\leq 1 $, the bottom-quark and tau distributions would look like the top-quark distribution, shown in the first row, third column.

By repeating a similar exercise with other parameters, we confirmed that the structure is primarily controlled by $\varepsilon$, with only a mild dependence on other parameters. The varying broadness of the distributions between $y_i$ for different generations is due to the fact that the first, second, and third family Yukawas are proportional to a product of three, two, and to a single UV coupling, respectively (see \cref{fig:diag}). Remember that their magnitudes are drawn from a uniform distribution in the $[0,1]$ range.

\begin{figure}[t]
    \centering
    \includegraphics[width=\linewidth]{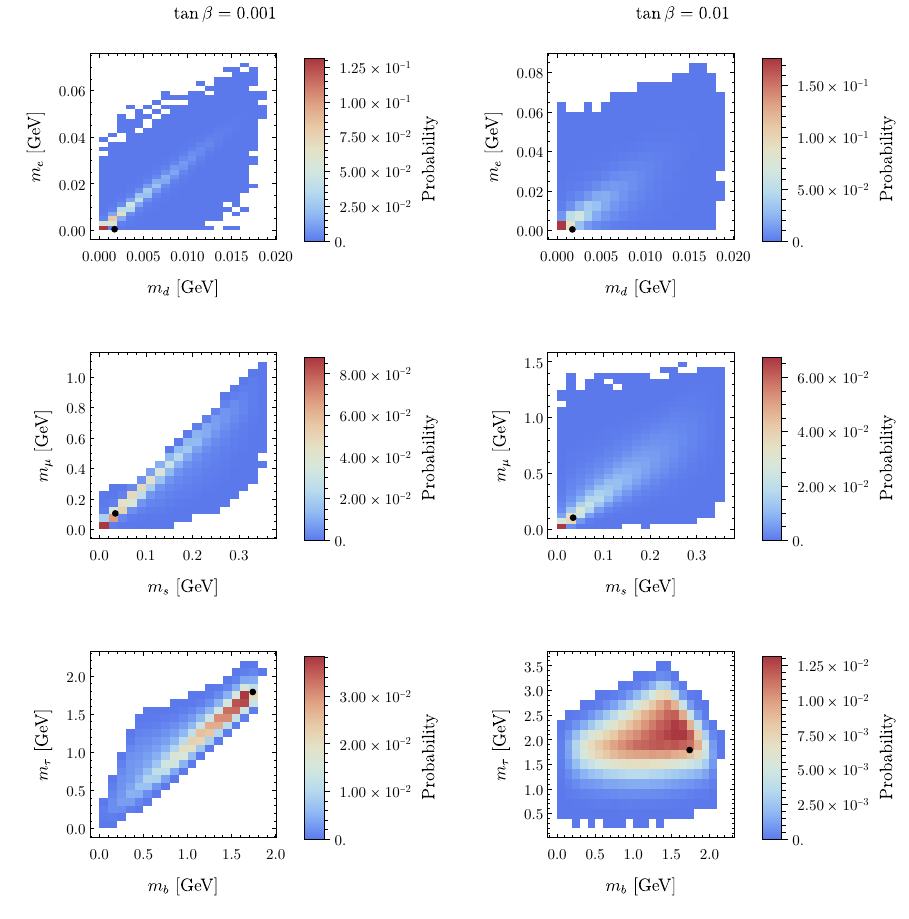}
    \caption{Two-dimensional histograms showing the correlation between the masses of the down quarks and charged leptons for two choices of $\tan \beta$. The black points mark the SM values at the chosen renormalization scale. See Section~\ref{sec:numerics2} for details.}
    \label{fig:dens}
\end{figure}

Another interesting aspect is revealed when studying the correlations between the observed parameters. To visualize the effect that $\beta$ has on the bottom-tau splitting, we generated one million parameter points for two different values of $\beta$: $0.001$ and $0.01$ while other configurations of the scan remain unchanged. In \cref{fig:dens}, we present histograms showing the correlation between the masses of the down-type quarks and charged leptons. The black dot marks the running masses at the appropriate scale. In the $\tan \beta \to 0$ limit, there is a strong correlation across all three generations, which is in remarkable agreement with observations! Increasing $\beta$ decreases the correlation, and this effect is most noticeable for the third generation.

%%%%%%
\subsection{Neutrino sector}
\label{sec:neutrino}
%%%%%%

Up to this point, we have focused on the flavor structure in the charged fermion sector, neglecting the neutrino masses and the Pontecorvo–Maki–Nakagawa–Sakata (PMNS)~\cite{Maki:1962mu} mixing matrix. Unlike the hierarchies observed in the charged sector, the neutrino flavor appears to be anarchic. Indeed, the PMNS matrix stands in stark contrast to the CKM matrix, exhibiting large mixing angles. Arguably, the only relevant hierarchy in the neutrino sector is related to the overall neutrino mass scale, which is many orders of magnitude below the electron mass. In this section, we will briefly discuss how to correctly reproduce neutrino parameters within this model.

At face value, the model predicts a Dirac mass matrix 
\begin{equation}\label{eq:Mnu}
M_\nu=\resizebox{.83\hsize}{!}{$\mqty(
 3b\qty(2\sqrt{6} v_1\lambda_{50}^*+v_2\lambda_\nu)Y_u^1 & 3b\qty(2\sqrt{6} v_1\lambda_{50}^*+v_2\lambda_\nu)Y_u^2 & 3b\qty(2\sqrt{6} v_1\lambda_{50}^*+v_2\lambda_\nu)Y_u^3 \\
 -12av_2 Y_u^1 & 3a\qty(2 \sqrt{6} v_1 y_u^2-3 v_2 Y_u^2) & 3a\qty(2 \sqrt{6} v_1 y_u^3-3 v_2 Y_u^3) \\
 -\sqrt{\frac{2}{3}} v_2 X_u^1 & -\sqrt{\frac{2}{3}} v_2 X_u^2 & v_1 x_u^3-\frac{1}{2} \sqrt{\frac{3}{2}} v_2 X_u^3)$}
\end{equation}
for the neutrinos, where
\begin{equation}
     \lambda_\nu=3\lambda_{45}+3\lambda_{46}- \lambda_{47}+2\lambda_{48}.
\end{equation}
This alone incorrectly implies that neutrino masses are similar in size to the up-quark masses, and the PMNS matrix is close to a unit matrix.

An elegant explanation of the smallness of neutrino masses in the context of minimal quark-lepton unification is provided by the \textit{inverse seesaw} mechanism~\cite{FileviezPerez:2013zmv}. Our model is minimally expanded by introducing three left-handed gauge-singlet fermions $S_i$ ($ i = 1, 2, 3 $). As a result, the Lagrangian gets supplemented by the following gauge-invariant and renormalizable operators:
\begin{equation}
    -\mathcal{L} \supset   \frac{1}{2} \mu_S^{ij} \bar S^c_i S_j +   Y^{ij}_R \bar S_i \chi^\dagger \psi_{u,\RR}^j + {\rm h.c.} \,,
\end{equation}
where $M^{ij}_R \equiv Y^{ij}_R \langle \chi \rangle$.
We define $ n_i = (\nu_{L i}, \nu_{R i}^{c}, S_i) $, where  $c$ denotes charge conjugation. The neutral lepton mass matrix, which is \( 9 \times 9 \), is given by
\begin{equation}
\mathcal{L}_{M_n} = - \frac{1}{2} \bar n_i M^{ij}_{n} n_j^{c} + {\rm h.c.}\,, \quad M_n = \begin{pmatrix}
0 & M_\nu & 0 \\
M_\nu^T & 0 & M_R^T \\
0 & M_R & \mu_S
\end{pmatrix},
\end{equation}
The global lepton number is an approximate symmetry softly broken by $\mu^{ij}_S$. For $ \mu_S \ll M_\nu < M_R $, the mass matrix for active neutrinos approximates to
\begin{equation}\label{eq:invseesaw}
m_\nu \approx M_\nu M_R^{-1} \mu_S (M_R^{-1})^T M^T_\nu~.    
\end{equation}
Thus, the overall smallness of the neutrino masses is explained by the small breaking of the lepton number via $\mu_S$. It is evident that the observed neutrino mass splittings and the PMNS mixing angles can be accurately fitted by introducing hierarchies into $ \mu_S $, given that $ M_\nu $ is hierarchical. The origin of these hierarchies is beyond the present work.

%%%%%%%%%%%%%%%%%%%%%%%
\section{Phenomenology} 
\label{sec:pheno}
%%%%%%%%%%%%%%%%%%%%%%

The symmetry-breaking scale of our model is constrained by rare processes mediated by the massive vector bosons resulting from this breaking. In Pati--Salam-type models, the vector leptoquarks are known to give contributions to the lepton flavor--violating decay $ K_L \to \mu e $, which is forbidden in the SM. This typically constrains the breaking scale to be of the order of a few PeV~\cite{Valencia:1994cj, Smirnov:2007hv}, and we will see that this general picture is reproduced here. The $ \SU(2)_{q+\ell} $ vector bosons are also known to give unsuppressed contributions to these SM-forbidden decays through a mechanism known as flavor transfer~\cite{Darme:2023nsy, Greljo:2023bix}, but the resulting bound on $ v_\Phi $ is weaker as we will show below. Typical rotations into the fermion mass basis will also result in contributions to muon conversion on heavy nuclei, a stringently constrained process with great potential for improvements in planned experiments.  

Integrating out the vector leptoquark and the flavored $ Z' $ at tree level results in effective operators 
    \begin{equation}
    \L_\sscript{SMEFT} \supset - \dfrac{1}{v_\chi^2} \Big| \overline{q}^3 \gamma_\mu \ell^3 + \overline{q}_\alpha \gamma_\mu \ell^\alpha + \overline{d}^p \gamma_\mu e^p \Big|^2 - \dfrac{1}{v_\Phi^2} \Big| \overline{q}_\alpha \gamma_\mu t^\alpha_{a\beta} q^\beta  + \overline{\ell}_\alpha \gamma_\mu t^\alpha_{a\beta} \ell^\beta \Big|^2
    \end{equation}
at the high scale, where $ t_a = \tfrac{1}{2} \sigma_a $ is the generator of $ \SU(2)_{q+\ell} $ rotations in the fundamental representation. We assume here that the right-handed neutrinos are too heavy to play a role in low-energy flavor physics and have ignored their contributions accordingly. Next, we transition to the low-energy effective Lagrangian (LEFT) below the scale of EW symmetry breaking to get to the mass basis fermions at the scale $ \SI{1}{GeV} $ relevant to the lepton flavor--violating processes discussed above. Only the semi-leptonic operators are relevant and shown here:\footnote{We have omitted a contribution to the left-handed vector operator proportional to $ [L^\dagger_e \mathcal{I} L_e]_{pr} $, as lepton flavor--violating contributions from this term are severely suppressed on account of a GIM-like mechanism.}
    \begin{align} \label{eq:LEFT_Lagrangian}
    \L_\sscript{LEFT} &\supset \,- \left( \dfrac{[L^\dagger_e L_d]_{pt} [L^\dagger_d L_e]_{sr} }{v_\chi^2} + \dfrac{[L^\dagger_e \mathcal{I} L_d]_{pt} [L^\dagger_d \mathcal{I} L_e]_{sr} }{v_\Phi^2} \right) (\overline{e}^p_\LL \gamma_\mu  e^r_\LL) (\overline{d}^s_\LL \gamma^\mu d^t_\LL) \nonumber \\
    & - \dfrac{[L^\dagger_e \mathcal{I} L_u]_{pt} [L^\dagger_u \mathcal{I} L_e]_{sr} }{v_\Phi^2} (\overline{e}^p_\LL \gamma_\mu  e^r_\LL) (\overline{u}^s_\LL \gamma^\mu  u^t_\LL) \\
    & + \bigg[ 6.0 \dfrac{[L^\dagger_e L_d]_{pt} [R^\dagger_d R_e]_{sr} }{v_\chi^2} (\overline{e}^p_\LL e_\RR^r) (\overline{d}^s_\RR d_\LL^t) \, + \, \mathrm{H.c.} \bigg]  - \dfrac{[R^\dagger_e R_d]_{pt} [R^\dagger_d R_e]_{sr} }{v_\chi^2} (\overline{e}^p_\RR \gamma_\mu  e^r_\RR) (\overline{d}^s_\RR \gamma^\mu  d^t_\RR ) \nonumber 
    \end{align}
where $ \mathcal{I} = \diag(1,\, 1,\,0) $. The numerical factor of $6.0$ on the second line is a combination of a factor of $ -2 $ from Fierzing the four-fermion operator and a factor of $ \sim 3.0 $ from running the operator down from $ \mu = \SI{3}{PeV} $ to $ \mu = \SI{1}{GeV} $ (as determined with \texttt{DsixTools}~\cite{Celis:2017hod,Fuentes-Martin:2020zaz}). 
By contrast, the vector current operators barely run at all (shifting only the operators by a few percent), and we disregard that effect. 

\subsection{Lepton flavor--violating kaon decay}

The non-observation of the $ K_L \to \mu e $ decay poses a strong constraint on the scale of symmetry breaking. The current experimental bound on the branching rate is $ \mathrm{Br}(K_L \to \mu e )<  \num{4.7e-12}$ @ 90\% CL~\cite{BNL:1998apv}. We calculated the NP contribution to this decay rate from the LEFT Lagrangian in \cref{eq:LEFT_Lagrangian} using Ref.~\cite{Marzocca:2021miv} and obtained
    \begin{align} \label{eq:KL_mue_contribution}
    \dfrac{\mathrm{Br}\big(K_L \to \mu e \big)}{\num{4.7e-12}} =\,&  \Big| 14\, \hat{v}^{\eminus 2}_\chi  \big( [L^\dagger_e L_d]_{ed} [R^\dagger_d R_e]_{s\mu} + [L^\dagger_e L_d]_{es} [R^\dagger_d R_e]_{d\mu} \big)^\ast \nonumber \\
    &\qquad - 0.10\, \hat{v}^{\eminus 2}_\chi \big( [L^\dagger_e L_d]_{\mu s} [L^\dagger_d L_e]_{de} + [L^\dagger_e L_d]_{\mu d} [L^\dagger_d L_e]_{se} \big) \nonumber\\
    &\qquad - 0.10\, \hat{v}^{\eminus 2}_\Phi \big( [L^\dagger_e \mathcal{I} L_d]_{\mu s} [L^\dagger_d \mathcal{I} L_e]_{de} + [L^\dagger_e \mathcal{I} L_d]_{\mu d} [L^\dagger_d \mathcal{I}L_e]_{se} \big)
    \Big|^2 \\
    &+ \Big| 14\, \hat{v}^{\eminus 2}_\chi \big( [L^\dagger_e L_d]_{\mu d} [R^\dagger_d R_e]_{se} + [L^\dagger_e L_d]_{\mu s} [R^\dagger_d R_e]_{de} \big) \nonumber\\
    &\qquad + 0.10\, \hat{v}^{\eminus 2}_\chi \big( [R^\dagger_e R_d]_{\mu s} [R^\dagger_d R_e]_{de} + [R^\dagger_e R_d]_{\mu d} [R^\dagger_d R_e]_{se} \big) \Big|^2, \nonumber
    \end{align}
having normalized the VEVs by $ \hat{v}_f \equiv v_f/ \SI{1}{PeV} $.
The pseudo-scalar quark current interpolates $ K_L $ better than the axial vector, resulting in a significant enhancement to the contributions stemming from mixed left--right currents due to the vector leptoquark, which are also enhanced by the RG running.\footnote{These are the contributions multiplying $ 14 \hat{v}^{\eminus 2}_\chi $.} 

With no suppression of the NP contribution from the fermion rotation matrices, the branching ratio in \cref{eq:KL_mue_contribution} bounds $ v_\chi \gtrsim \SI{4}{PeV} $. However, it may be possible to evade this constraint. In a simple approximation, where $ L_e^\dagger L_d $ and $ R_d^\dagger R_e $ are taken to be two-dimensional rotation matrices (dominated by the 1--2 rotation angle), the scalar contribution will vanish altogether if the rotation angles are orthogonal. While the left-handed rotations are perturbative and, therefore, close to the identity matrix, the right-handed rotation might be close to maximal, leading to a vanishing scalar contribution. In this case, the bound will come from the vector currents constraining $ v_\chi, v_\Phi \gtrsim \SI{300}{TeV} $. That being said, we consider the realizations with the symmetry-breaking scale of at least a PeV to be the more realistic scenario.

\subsection{Muon conversion on heavy nuclei}
Muon conversion on heavy nuclei is closely correlated to the lepton flavor--violating kaon decay, which is also caused by the Lagrangian in \cref{eq:LEFT_Lagrangian}, but without any quark flavor violation. The SINDRUM-II experiment set the best present bound on this process, obtaining the limit $ \mathrm{Cr}(\mu \, \mathrm{Au} \to e \, \mathrm{Au}) < \num{7e-13}$ for the conversion rate in muonic gold~\cite{SINDRUMII:2006dvw}.
Scalar and vector four-fermion operators contribute to this process at roughly the same rate, and, using Ref.~\cite{Kitano:2002mt}, we determine the contribution from the heavy vector bosons:
    \begin{align} \label{eq:muon_conversion}
    \dfrac{\mathrm{Cr}(\mu \, \mathrm{Au} \to e \, \mathrm{Au})}{\num{7e-13}} = \,& \Big| 0.5 \,\hat{v}^{\eminus 2}_\Phi [L^\dagger_e \mathcal{I} L_u]_{e u} [L^\dagger_u \mathcal{I} L_e]_{u\mu}
    + 0.6 \,\hat{v}^{\eminus 2}_\chi [L^\dagger_e L_d]_{e d} [L^\dagger_d L_e]_{d\mu} \\
    & \hspace{-2cm} + 0.6 \,\hat{v}^{\eminus 2}_\Phi [L^\dagger_e \mathcal{I} L_d]_{e d} [L^\dagger_d \mathcal{I} L_e]_{d\mu}
    + 2.7 \hat{v}_\chi^{\eminus 2} [R^\dagger_e R_d]_{e d} [L^\dagger_d L_e]_{d\mu} 
    + 1.4 \hat{v}_\chi^{\eminus 2} [R^\dagger_e R_d]_{e s} [L^\dagger_d L_e]_{s\mu}
    \Big|^2  \nonumber \\
    & \hspace{-2cm}+ \Big| 
    0.6 \,\hat{v}^{\eminus 2}_\chi [R^\dagger_e R_d]_{e d} [R^\dagger_d R_e]_{d\mu} 
    + 2.7 \hat{v}_\chi^{\eminus 2} [L^\dagger_e L_d]_{e d} [R^\dagger_d R_e]_{d\mu} 
    + 1.4 \hat{v}_\chi^{\eminus 2} [L^\dagger_e L_d]_{e s} [R^\dagger_d R_e]_{s\mu}
    \Big|^2  \nonumber 
    \end{align}
The numerical enhancement of the scalar operators is due to the RG-running already present in the Wilson coefficients in \cref{eq:LEFT_Lagrangian}. 

Muon conversion violates lepton flavor by one unit while quark flavor is conserved. Thus, in the interaction basis, the process would be forbidden, as no flavor transfer~\cite{Darme:2023nsy} is possible. We should, however, expect a mixing of at least the left-handed down-type in order to recover the sizable Cabbibo angle. In the limit, where the $ [L_d]_{12} = \theta_c = 0.22 $ is the only non-trivial mixing, the NP contribution in \cref{eq:muon_conversion} gives the bounds $ v_\chi \gtrsim \SI{0.8}{PeV}$ and $ v_\Phi \gtrsim \SI{0.4}{PeV}$. While specific realizations of the mixing matrices may either enhance or weaken the bounds, they are a good estimate for realistic benchmarks in the absence of tuning. Interestingly, the future Mu2e~\cite{Bernstein:2019fyh} and COMET~\cite{Moritsu:2022lem} experiments projects an improvement on the experimental sensitivity to the level of $ \num{e-17} $.\footnote{Both Mu2e and COMET will be using aluminum as the target nuclei, changing the numerical factors in Eq.~\eqref{eq:muon_conversion}.} Hence, the experimental sensitivity to the symmetry-breaking scales is expected to improve by an order of magnitude. This provides an exciting avenue for experimentally detecting the imprints of the model behind the flavor structure posited in this paper.

%%%%%%%%%%
\section{Conclusion}
\label{sec:conc}
%%%%%%%%%%

The field of flavor physics has a bright future due to a comprehensive experimental program anticipated within this decade. The interest is driven by the potential to detect indirect effects of heavy BSM physics from scales well beyond the reach of direct searches. Indirect signs of new physics might manifest in phenomena such as rare flavor-changing neutral currents, violations of charged lepton flavor, and electric dipole moments, each of which is expected to see substantial experimental advancements~\cite{LHCb:2018roe, LHCb:2021glh, Belle-II:2018jsg, Forti:2022mti, Belle-II:2022cgf,  NA62KLEVER:2022nea, MEGII:2018kmf, Bernstein:2019fyh, Moritsu:2022lem, n2EDM:2021yah, Wu:2019jxj, Blondel:2013ia, ACME:2018yjb, EuropeanStrategyforParticlePhysicsPreparatoryGroup:2019qin}. Furthermore, flavor physics is closely tied to some of the most significant structural challenges of the SM. The origin of flavor remains an enduring enigma: one of the most puzzling aspects is the presence of hierarchies in the masses (and mixings) of charged fermions between successive generations, despite all of them arising from the Yukawa interactions with a single Higgs field. This longstanding puzzle has attracted a renewed interest in light of recent experimental activities. 

Perhaps the resolution of the flavor puzzle may lie already at the next hidden layer of nature. It is worthwhile to explore novel possibilities for low-scale flavor models, which could lead to deviations detectable in forthcoming experiments. In this paper, we construct and thoroughly investigate one particularly elegant example. Building upon our recent work~\cite{Greljo:2023bix}, we combine a gauged $\SU(2)$ flavor symmetry with a minimal quark-lepton unification paradigm based on the $\SU(4)$ gauge group. The flavor symmetry acts on left-chiral fermions, distinguishing the light-family doublet from the third-family singlet. The model, detailed in \cref{sec:model}, predicts hierarchically distinct contributions to the rank of the Yukawa matrices, which then translates into the hierarchies between families. 

As illustrated in \cref{fig:diag}, the third row in the Yukawa matrices results from unsuppressed dimension-4 Yukawa interactions. The second row emerges from integrating out a single heavy vector-like fermion, yielding a tree-level dimension-5 operator. The first row is populated by loop diagrams involving the same fields and generating a loop-level dimension-5 operator. The matter field content required to realize this mechanism is remarkably minimal, as shown in \cref{tab:matter_fields}. This is due to the elegant unification of multiple elements from Ref.~\cite{Greljo:2023bix} collected in \cref{tab:component_fields}. 

The entire SM flavor structure is controlled by a single parameter $\varepsilon = v_\Phi/M_\Psi$, a ratio of the flavor-breaking scale to the VLF mass. After performing a detailed matching calculation to the SM Yukawas in \cref{sec:matching}, we conduct a comprehensive numerical scan in \cref{sec:numerics}. The key result is presented in \cref{fig:hist}, showing a histogram of predictions for SM flavor parameters, for $\varepsilon = 10^{-2}$, obtained from randomly selecting dimensionless UV couplings from a flat distribution in the range $|y_x| \leq 1$. Remarkably, reasonable order-of-magnitude predictions emerge for all measured parameters, except for the $b/\tau$ Yukawa, which requires a single UV parameter to be somewhat smaller than expected, $x_d^3 \sim 10^{-2}$. 
Even more remarkable is the predicted correlation between down-quark and charged lepton masses shown in \cref{fig:dens}. For a small ratio of Higgs VEVs, $\tan\beta \to 0$, this correlation is very precise and shows an extraordinary agreement with the measured values. This exemplifies how quark-lepton unification, a well-motivated concept on its own, aids in explaining the flavor puzzle.

As demonstrated in \cref{sec:pheno}, the leading signatures of the model are lepton flavor--violating kaon decays and muon conversion on heavy nuclei, where the experimental prospects are excellent. In fact, current limits already require the scale of the model to be at least of the order of PeV, which introduces another hierarchy problem related to the smallness of the electroweak scale. Additionally, generating small neutrino masses can be straightforwardly achieved using the inverse see-saw mechanism, as described in \cref{sec:neutrino}. However, predicting large PMNS mixing angles requires additional structure. These issues are not addressed in this work and are left for future research. 

\subsection*{Acknowledgments}

This work has received funding from the Swiss National Science Foundation (SNF) through the Eccellenza Professorial Fellowship ``Flavor Physics at the High Energy Frontier,'' project number 186866, and the Ambizione grant ``Matching and Running: Improved Precision in the Hunt for New Physics,'' project number 209042.

\appendix
\section{Scalar potential}\label{app:A}
To find the full scalar potential and ensure that no superfluous terms were included or any possible terms omitted, we made use of the \textsc{Mathematica} package \texttt{Sym2Int} \cite{Fonseca:2017lem}. We have used Roman letters for $\SU(2)_\LL$ indices: $i,j$, and Greek letters for $\SU(2)_{q+\ell}$ indices: $\alpha,\beta,\gamma,\delta$, and surpressed all other indices. The full scalar potential is given below:
\begin{align}
    V&=m_{\chi}^2\abs{\chi}^2+m_{H_1}^2\abs{H_1}^2+2m_{\Sigma_H}^2\Tr\!\big(\abs{\Sigma_H}^2)+2m_{\Sigma_\Phi}^2\Tr\!\big(\abs{\Sigma_\Phi}^2)\nonumber\\ &+\lambda_1\abs{\chi}^4+\lambda_2\abs{H_1}^2\abs{\chi}^2+\lambda_3\abs{H_1}^2\Tr\!\big(\abs{\Sigma_H}^2)+\lambda_4\abs{\chi}^2\Tr\!\big(\abs{\Sigma_H}^2)\nonumber\\ &+(\lambda_5 H_{1i}^\dagger\chi^\dagger\Sigma_H^i\chi+\text{h.c.})+\lambda_6 H_{1i}^\dagger\Tr\!\big(\Sigma_{Hj}^\dagger\Sigma_H^i)H^j+\lambda_7\chi^\dagger\Sigma_H^i\Sigma_{Hi}^\dagger\chi+\lambda_8\chi^\dagger\abs{\Sigma_H}^2\chi\nonumber\\ &+\lambda_9\abs{H_1}^4+\lambda_{10}\Tr\!\big(\abs{\Sigma_H}^4)+\lambda_{11}\qty[\Tr\!\big(\abs{\Sigma_H}^2)]^2+\qty(\lambda_{12}H_{1i}^\dagger\Tr\!\big(\Sigma_H^i\Sigma_H^j)H_{1j}^\dagger+\text{h.c.})\nonumber\\ &+\qty(\lambda_{13}H_{1i}^\dagger\Tr\!\big(\Sigma_H^i\abs{\Sigma_H}^2)+\lambda_{14}H_{1i}^\dagger\Tr\!\big(\Sigma_H^i\Sigma_H^j\Sigma_{Hj}^\dagger)+\text{h.c.})+\lambda_{15}\Tr\!\big(\Sigma_{Hi}^\dagger\Sigma_H^j\Sigma_{Hj}^\dagger\Sigma_H^i)\nonumber\\ &+\lambda_{16}\Tr\!\big(\Sigma_{Hi}^\dagger\Sigma_H^j)\Tr\!\big(\Sigma_{Hj}^\dagger\Sigma_H^i)+\lambda_{17}\Tr\!\big(\Sigma_{Hi}^\dagger\Sigma_{Hj}^\dagger)\Tr\!\big(\Sigma_H^i\Sigma_H^j)\nonumber\\
    &+\lambda_{18}\Tr\!\big(\Sigma_{Hi}^\dagger\Sigma_{Hj}^\dagger\Sigma_H^i\Sigma_H^j)+\lambda_{19}\Tr\!\big(\Sigma_{Hi}^\dagger\abs{\Sigma_H}^2\Sigma_H^i)+\lambda_{20}\abs{H_1}^2\Tr\!\big(\abs{\Sigma_\Phi}^2)\nonumber\\
    &+\lambda_{21}\abs{\chi}^2\Tr\!\big(\abs{\Sigma_\Phi}^2)+\lambda_{22}\chi^\dagger\abs{\Sigma_\Phi}^2\chi+\lambda_{23}\chi^\dagger\Sigma_\Phi^\alpha\Sigma_{\Phi\alpha}^\dagger\chi+\lambda_{24}\epsilon_{\alpha\beta}\chi^\dagger\Sigma_\Phi^\alpha\Sigma_\Phi^\beta\chi\nonumber\\ &+\lambda_{25}\Tr\!\big(\abs{\Sigma_\Phi}^4)+\lambda_{26}\qty[\Tr\!\big(\abs{\Sigma_\Phi}^2)]^2+\lambda_{27}\Tr\!\big(\Sigma_{\Phi\alpha}^\dagger\Sigma_\Phi^\beta)\Tr\!\big(\Sigma_{\Phi\beta}^\dagger\Sigma_\Phi^\alpha)\nonumber\\ &+\lambda_{28}\Tr\!\big(\Sigma_{\Phi\alpha}^\dagger\Sigma_{\Phi\beta}^\dagger)\Tr\!\big(\Sigma_\Phi^\alpha\Sigma_\Phi^\beta)+\lambda_{29}\Tr\!\big(\Sigma_{\Phi\alpha}^\dagger\Sigma_\Phi^\beta\Sigma_{\Phi\beta}^\dagger\Sigma_\Phi^\alpha)\nonumber\\
    &+\lambda_{30}\Tr\!\big(\Sigma_{\Phi\alpha}^\dagger\Sigma_{\Phi\beta}^\dagger\Sigma_\Phi^\alpha\Sigma_\Phi^\beta)+\lambda_{31}\Tr\!\big(\Sigma_{\Phi\alpha}^\dagger\abs{\Sigma_\Phi}^2\Sigma_\Phi^\alpha)\nonumber\\
    &+\qty(\lambda_{32}\epsilon_{\alpha\beta}\epsilon_{\gamma\delta}\Tr\!\big(\Sigma_\Phi^\alpha\Sigma_\Phi^\gamma)\Tr\!\big(\Sigma_\Phi^\beta\Sigma_\Phi^\delta)+\lambda_{33}\epsilon_{\alpha\beta}\epsilon_{\gamma\delta}\Tr\!\big(\Sigma_\Phi^\alpha\Sigma_\Phi^\gamma\Sigma_\Phi^\beta\Sigma_\Phi^\delta)+\text{h.c.})\nonumber\\
    &+\qty(\lambda_{34}\epsilon_{\alpha\beta}\Tr\!\big(\Sigma_\Phi^\alpha\Sigma_{\Phi\gamma}^\dagger)\Tr\!\big(\Sigma_\Phi^\beta\Sigma_\Phi^\gamma)+\lambda_{35}\epsilon_{\alpha\beta}\Tr\!\big(\Sigma_\Phi^\alpha\abs{\Sigma_\Phi}^2\Sigma_\Phi^\beta)+\text{h.c.})\nonumber\\
    &+\qty(\lambda_{36}\epsilon_{\alpha\beta}\Tr\!\big(\Sigma_\Phi^\alpha\Sigma_\Phi^\gamma\Sigma_{\Phi\gamma}^\dagger\Sigma_\Phi^\beta)+\text{h.c.})+\lambda_{37}\Tr\!\big(\abs{\Sigma_H}^2)\Tr\!\big(\abs{\Sigma_\Phi}^2)\nonumber\\
    &+\lambda_{38}\Tr\!\big(\Sigma_{Hi}^\dagger\Sigma_\Phi^\alpha)\Tr\!\big(\Sigma_{\Phi\alpha}^\dagger\Sigma_H^i)+\lambda_{39}\Tr\!\big(\Sigma_{Hi}^\dagger\Sigma_{\Phi\alpha}^\dagger)\Tr\!\big(\Sigma_H^i\Sigma_\Phi^\alpha)+\lambda_{40}\Tr\!\big(\abs{\Sigma_H}^2\abs{\Sigma_\Phi}^2)\nonumber\\
    &+\lambda_{41}\Tr\!\big(\Sigma_{Hi}^\dagger\Sigma_{\Phi\alpha}^\dagger\Sigma_H^i\Sigma_\Phi^\alpha)+\lambda_{42}\Tr\!\big(\Sigma_{Hi}^\dagger\abs{\Sigma_\Phi}^2\Sigma_H^i)+\lambda_{43}\Tr\!\big(\Sigma_{\Phi\alpha}^\dagger\abs{\Sigma_H}^2\Sigma_\Phi^\alpha)\nonumber\\
    &+\lambda_{44}\Tr\!\big(\Sigma_{Hi}^\dagger\Sigma_\Phi^\alpha\Sigma_H^i\Sigma_{\Phi\alpha}^\dagger)+\lambda_{45}\Tr\!\big(\Sigma_H^i\Sigma_{Hi}^\dagger\Sigma_\Phi^\alpha\Sigma_{\Phi\alpha}^\dagger)\nonumber\\
    &+\qty(\lambda_{46}\epsilon_{\alpha\beta}\Tr\!\big(\Sigma_{Hi}^\dagger\Sigma_\Phi^\alpha)\Tr\!\big(\Sigma_H^i\Sigma_\Phi^\beta)+\lambda_{47}\epsilon_{\alpha\beta}\Tr\!\big(\Sigma_\Phi^\alpha\abs{\Sigma_H}^2\Sigma_\Phi^\beta)+\text{h.c.})\nonumber\\
    &+\qty(\lambda_{48}\epsilon_{\alpha\beta}\Tr\!\big(\Sigma_\Phi^\alpha\Sigma_H^i\Sigma_{Hi}^\dagger\Sigma_\Phi^\beta)+\lambda_{49}\epsilon_{\alpha\beta}\Tr\!\big(\Sigma_\Phi^\alpha\Sigma_{Hi}^\dagger\Sigma_\Phi^\beta\Sigma_H^i)+\text{h.c.})\nonumber\\
    &+\qty(\lambda_{50}\epsilon_{\alpha\beta}H_{1i}^\dagger\Tr\!\big(\Sigma_H^i\Sigma_\Phi^\alpha\Sigma_\Phi^\beta)+\lambda_{51}\epsilon_{\alpha\beta}H_1^i\Tr\!\big(\Sigma_{Hi}^\dagger\Sigma_\Phi^\alpha\Sigma_\Phi^\beta)+\text{h.c.})\nonumber\\
    &+\qty(\lambda_{52}H_{1i}^\dagger\Tr\!\big(\Sigma_H^i\abs{\Sigma_\Phi}^2)+\lambda_{53}H_{1i}^\dagger\Tr\!\big(\Sigma_H^i\Sigma_\Phi^\alpha\Sigma_{\Phi\alpha}^\dagger)+\text{h.c.})~.
\end{align}

\bibliographystyle{JHEP}
\bibliography{bibliography}

\end{document}